\documentclass[aps,pre,showpacs,twocolumn]{revtex4}
\usepackage{amssymb}
\usepackage{graphicx}
\usepackage{colordvi}
\usepackage{color}
\usepackage{dcolumn,amsmath,amsthm,amscd,amsfonts,amssymb,epsfig,graphics,graphicx,eucal}

\newcommand{\sn}{\mbox{sn}}

\begin{document}

\title{Superfluidity breakdown  of periodic matter waves in  quasi one-dimensional annular traps via resonant scattering  with moving defects}
\author{A.V. Yulin$^{1}$, Yu. V. Bludov$^2$, V. V. Konotop$^{1}$, V. Kuzmiak$^{3}$, and M. Salerno $^4$}
\affiliation{ $^1$Centro de F\'isica Te\'orica e Computacional and Departamento de F\'isica, Faculdade de Ci\^encias,
Universidade de Lisboa,   Avenida Professor
Gama Pinto 2, Lisboa 1649-003, Portugal
\\
$^2$  Centro de F\'{\i}sica, Universidade do Minho, Campus de Gualtar, Braga 4710-057, Portugal
\\
$^3$ Institute for Photonics and Electronic, Czech Academy of Science, Chaberska 57, 182 51, Praha 8, Czech Republic
\\
$^4$Dipartimento di Fisica ``E.R. Caianiello'', Universit\`a di Salerno, Via Ponte don Melillo, 84084 Fisciano (SA), Italy}

\date{\today}

\begin{abstract}
We investigate, both analytically and numerically, the quasi-superfluidity properties of periodic Bose-Einstein condensates (BECs) in a quasi-one-dimensional (1D) ring with  optical lattices (OL) of different kinds (linear and nonlinear) and with a moving  defect of an infinite mass inside. To study the dynamics of the condensate we used a mean-field approximation describing the condensate by use of the Gross-Pitaevskii equation for the order parameter. We show that the resonant scattering of sound Bloch waves with the defect profoundly affect  BEC superfluidity. In particular, a  moving defect always leads to the breakdown of superfluidity independently of the value of its velocity. For weak periodic potentials the superfluidity breakdown may occur on a very long time scale (quasisuperfluidity)  but the breakdown process can be accelerated by increasing the strength of the OL. Quite remarkably, we find that when the length of the ring is small enough to imply the discreteness of the reciprocal space, it becomes possible to avoid the resonant scattering and to restore quasi-superfluidity.
\end{abstract}
\maketitle

\input{epsf.tex}
\epsfverbosetrue

\section{Introduction}
\label{intro}
As is well known, superfluidity is  the remarkable  property of quantum fluids to flow without any dissipation \cite{Leggett}.  A criterion for this phenomenon to occur,  first proposed by Landau \cite{Landau}, is the existence of a critical velocity below which no excitation can be created in the fluid, obviously implying frictionless motion for any velocity below the critical one.  Existence of the critical velocity in liquid $^4$He has been experimentally confirmed, although for much lower values than the theoretical  predictions. This  discrepancy was later explained  by Feynman \cite{Feynman} in terms of vortex ring formation as an alternate mechanism for dissipation in the fluid.

A similar situation  has been found in novel quantum fluids realized as  Bose-Einstein condensate (BEC) gases. The extreme flexibility of these systems in  interactions and parameters design makes them ideal for superfluidity investigations. In particular, the interaction of a BEC with a moving defect is taken as a  test bed for the gas superfluidity. Thus, a defect moving through a BEC with a velocity below a critical velocity  (Landau critical velocity coincides in this case with the sound velocity) is expected to experience no drag force while a nonzero drag force should appear in the supersonic regime, due to Cherenkov radiation~\cite{AstPit}. The existence of the critical velocity in BECs has been experimentally observed in the MIT experiment \cite{MIT} for lower (undersonic) values than what was expected, due also in this case to vortex formation. Recently, the phenomena of sub- and supersonic motions of defects in polariton BECs have been reported \cite{BEC_polariton_t}, \cite{BEC_polariton_exp}. It was shown there both theoretically and experimentally that the motion of the condensate is superfluid until the velocity of the defect exceeds some critical velocity.

Superfluid currents through a set of equally spaced impurities (barriers) modeling an optical lattice (OL) have also been reported~\cite{lattice}. The main feature introduced by the periodic potential is  that the superfluid and sound waves are Bloch functions and the dispersion relation  differs from the usual  dispersion relation of Bogoliubov phonons for homogeneous condensates \cite{bogoliubov}. In a 1D setting the BEC superfluidity in OLs can be  lost via the energetic instability
as well as  through dynamical instabilities or  modulational instabilities~\cite{KS,instability} of Bloch waves  leading to soliton formation. This last instability can be viewed as the 1D analog of the vortex formation in multidimensional superfluid settings. Another possibility for the breakdown of superfluidity  is the occurrence of superfluid-insulator transitions~\cite{transitions} in the limit of very  deep  OLs. The breakdown of superfluidity of a trapped condensate moving in a 1D OL has been experimentally reported in Ref.~\cite{Burger}.

It is also known that the critical velocity may be sensitive both to the dimension and to the geometry of the trap. In this  context persistent currents in annular traps attract particular attention, having been addressed both theoretically~\cite{annular_theory} and experimentally~\cite{annular_experiment}.
One may expect from this that the introduction of  a moving  defect in a periodic BEC  can manifest profoundly different responses of the superfluid due to the interplay of periodicity and nonlinearity which could induce (enhance)  resonances  between sound waves and the moving defect. Moreover, the superfluid response could depend on whether the defect is at rest with respect to the OL  (e.g., superfluid BEC flow though an OL and a defect both stationary)  or  if it is moving with respect to the OL. The response could depend as well on the shape and strength of the defect since the appearance of nonlinear defect modes is also possible~\cite{defects, bs11}.

The aim of the present paper is to study superfluid properties of periodic BECs moving in quasi-1D ring traps in the presence of OLs of different kinds, e.g., linear and nonlinear,  and  with a localized defect moving with respect to them. We note that persistent BEC currents in  quasi-1D rings are possible both in the presence of a standard OL (e.g., a linear periodic  potential) and in the presence of a periodic modulation of the scattering length along the  ring (also called nonlinear OL~\cite{nonlin_OL}),  as recently discussed  in Ref.~\cite{YBKKS}.
In this context we show, as a result,  the occurrence of
a dynamical instability which arises  from resonances in the defect-sound waves dynamics (resonant scattering) and always leading to the breakdown of BEC superfluidity,  independently  of defect velocity. This instability, although of dynamical type, should not not be confused  with the usual  dynamical instability reported for periodic BEC systems (e.g., modulational instability of Bloch waves)~\cite{KS}. Let us mention here that this instability can be also treated as transitional radiation \cite{Landau_add} or as Cherenkov radiation of Bloch waves \cite{Ch_b1}-\cite{Ch_b7}. For weak periodic potentials the loss of BEC superfluidity via resonant scattering may occur on a time scale so  long that can be considered infinite for any practical purpose. The superfluidity breakdown, however, can be clearly detected because it can be accelerated by increasing the strength of the OL. Quite remarkably, we find that when the size of the ring in which the BEC is confined is small enough to imply the discreteness of the reciprocal (crystal momentum) space, it becomes possible to avoid resonances with a proper design of the system parameters and allows us to achieve quasi-superfluidity or full superfluidity behavior.

The  paper is organized as follows. In Sec. II we present the model equation and discuss families of current-carrying states for different settings. In Sec. III we study the Bogoliubov spectrum of the Bloch sound waves propagating against periodic BEC current-carrying backgrounds. In Sec.  IV we use perturbation theory to give a quantitative picture of the sound dynamics induced by a defect moving in a periodic BEC in a ring and to discuss the resonant conditions for the scattering between the defect and sound excitations
to occur. Section V is devoted to the study of the BEC superfluidity  in the case of a  periodic quasi-infinite geometry setting in the presence of a linear or a nonlinear OL. In particular, we use direct numerical integrations of GPE to show that,  in agreement with our resonant condition analysis, the BEC superfluidity may always be lost  by resonant scattering, although the real breakdown may occur on a very long time scale. In Sec. VI we address the same problem for the case of a finite-size ring geometry  and show that when the ring is small enough  it becomes  possible to  avoid the scattering resonances and to restore superfluidity. Finally, in Sec. VII, the main results of the paper are briefly summarized.

\section{Model equation}
\label{sec:model}

We consider the  one-dimensional Gross-Pitaevskii equation (GPE)
\begin{eqnarray}
\label{GP}
i \psi_t = -\psi_{xx} + V(x)\psi + U(x)|\psi|^2\psi + V_d (x-v_d t) \psi,
\end{eqnarray}
with $V(x)$ and $U(x)$ as the periodic linear and nonlinear potentials, respectively, which account for external optical or magnetic traps (lattices) and for space-varying interatomic interactions, and with   $V_d(x-v_d t)$ a localized linear potential modeling a defect moving with the velocity $v_d$ and  perturbing a given superfluid flow.   Without loss of generality,  the period of the lattices is chosen to be $\pi$, i.e., $V(x+\pi)=V(x)$, $U(x+\pi)=U(x)$. This scaling implies that the energy is measured in the units of the recoil energy. We assume that the length of the ring (annular) trap is $L=n \pi$, where $n$ is an integer, and   focus on the case when $n\gg 1$. This geometry implies cyclic boundary conditions,
\begin{equation}
\label{cyclic}
\psi(x)=\psi(x+L).
\end{equation}
In particular, we are interested in nonzero current states, which in the absence of the defect (i.e., when $V_d (x-v_d t)\equiv 0$),  can be represented in the form of the nonlinear Bloch state   $\Psi_{v_0}=e^{-i\mu t+iv_0 x}\psi_{v_0}(x)$ with the mean superfluid velocity $v_0$ playing the role of the crystal momentum and with $\psi_{v_0}(x)$ a complex periodic solution of the equation  (see also Ref.\cite{YBKKS})
\begin{figure}[h]
\epsfig{file=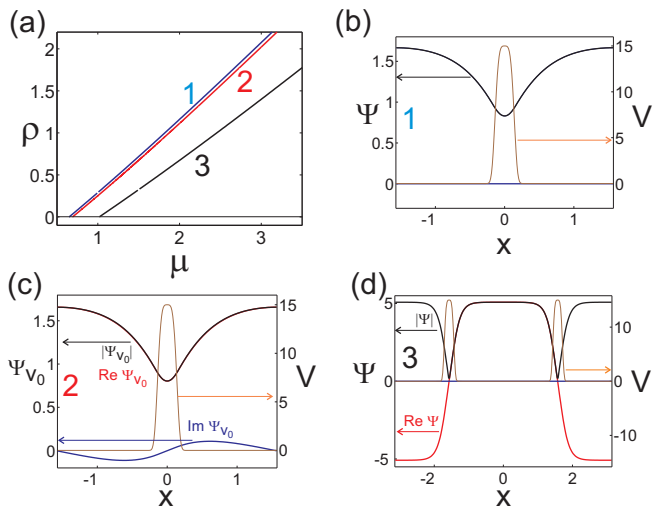,angle=0,width=\columnwidth}
\caption{(Color online) (a) Families of the stationary backgrounds with the
velocity $v_0=0$ (blue line, 1),  $v_0=0.25$ (red line, 2) and $v_0=1$ (black line, 3) in the presence of a linear OL only [the nonlinear potential is homogeneous $U(x)=1$].  Examples of stationary solutions for $\mu=3$ of Eq.~(\ref{stationary}) are shown in panels (b)-(d). Panel (b) marked also by big blue number 1, shows the field distribution corresponding to the bifurcation curve 1 in panel (a). The left vertical axis of panel (b) is for the field $\psi$ (which is pure real in this case) and the right vertical axis is for linear potential $V$.
Panel (c) shows the distribution of the absolute value, real and imaginary parts of the complex field $\psi_{v_0}$ which correspond to a nonzero current state [bifurcation curve 2 in panel (a)].
Panel (d) is the same but for bifurcation curve 3 in panel (a). Here the left vertical axis shows the absolute value and the real part of the field $\psi$, the imaginary part of $\psi$ is zero because this distribution corresponds to a currentless state.
The parameters are
 $V_0=15$, $\ell =0.15$, $x_s=0$ [panels (b) and (d)], or $x_s=\pi/2$ [panel (c)].}
\label{fig1_1}
\end{figure}
\begin{eqnarray}
\label{stationary}
&& \mu \psi_{v_0}= - \psi_{v_0,xx} - 2 i v_0 \psi_{v_0,x}  \nonumber  \\
&& \qquad\qquad + \left[v_0^2 + V(x) +  U(x)|\psi_{v_0}|^2\right]\psi_{v_0}
\end{eqnarray}
 with $\mu$ denoting  the chemical potential.
For the sake of definiteness, below we explore the linear potential $V$ of a super-Gaussian form
\begin{eqnarray}
\label{potentialV}
V(x)=V_0\sum_m e^{-\frac{(x-x_m)^4}{\ell^4}}
\end{eqnarray}
whose peaks have the characteristic widths $\ell$ and depths $V_0$, and are centered in the points $x_m=\pi m+x_s$,  with the shift $x_s$ introduced for the sake of convenience.

The nonlinear lattice is modeled by the potential of the form
\begin{eqnarray}
\label{potentialU}
U(x)=\frac{\kappa q^2}{4}\frac{\sn(q[x-\pi/4],\kappa )(3\kappa\,\sn(q[x-\pi/4], \kappa)-2)}{1+\kappa \sn(q[x-\pi/4],\kappa)}, \\ q=\frac{4K(\kappa)}{\pi},\nonumber
\end{eqnarray}
where we use the standard notations for the Jacobi elliptic function $\sn$ and for the elliptic integral $K(\kappa)$.
\begin{figure}[h]
\epsfig{file=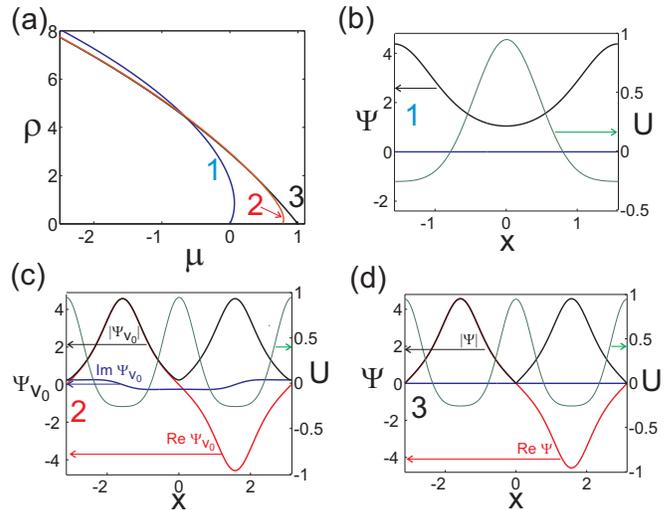,angle=0,width=\columnwidth}
\caption{(Color online) (a) Families of the stationary backgrounds with the
velocity $v_0=1$ (blue line, 1),  $v_0=0.125$ (red line, 2) and $v_0=0$ (black line, 3) in the presence of only nonlinear OL [linear OL $V(x)\equiv 0$].  Examples of stationary solutions for $\mu=-2$ of Eq.~(\ref{stationary}) are shown in panels (b)-(d). Panel (b), marked also by big blue number 1 shows the field distribution corresponding to the bifurcation curve 1 in panel (a). The left vertical axis of panel (b) is for the field $\psi$ (which is pure real in this case) and the right vertical axis is for the nonlinear potential $U$.
Panel (c) shows the distribution of the absolute value, real and imaginary parts of the complex field $\psi_{v_0}$ corresponding to the bifurcation curve 2 in panel (a).
Panel (d) is the same but for bifurcation curve 3 in panel (a). Here the left vertical axis shows the absolute value and the real part of the field $\psi$, the imaginary part of $\psi$ is zero.
 The nonlinear potential $U(x)$ (5) is centered at $x = \pi/4$ with $\kappa = 0.25$  likewise in Ref. \cite{YBKKS}.}
\label{fig1_2}
\end{figure}
Obviously $U(x)$  is parametrized by the elliptic modulus  $\kappa \in[0,1]$.  This form of the potential, centered at $x = \pi/4$ with $\kappa = 0.25$,  is the same as the one used in Ref.\cite{YBKKS} to study superfluid BEC ground states in a nonlinear OL.

Notice that while the linear and nonlinear  potentials are taken with  period $\pi$ the complex function $\psi_{v_0}(x)$ may have  periods $L_0 = m\pi$ with $m$ an integer. In the following, we mainly focus on  $\pi$-periodic solutions for the case of linear OL and on $2\pi$-periodic solutions for the case of nonlinear OL. The reason for choosing these solutions is because they are both stable for the respective cases (the $\pi$-periodic solutions are always stable for linear OLs while for nonlinear OLs  the $2\pi$-periodic solutions are stable for large strongly varying densities of condensate, e.g., when the  condensate is in the form of an interacting chain of droplets, with relatively large values of the chemical potential).

In Fig.~\ref{fig1_1}(a) we present examples of the families of stationary periodic solutions on the plane $(\rho,\mu)$, where $\rho=\frac 1L\int_{0}^{L}|\Psi (x)|^2 dx$ is the average density of the condensate, as well as, examples of the profiles of the density distribution in the presence of linear lattice only [Figs.~\ref{fig1_1}(b)--\ref{fig1_1}(b)(d)].
One can see that the density of the condensate varies significantly; therefore, one can expect that interaction with upper branches of the dispersion spectra of the linear excitations may be important.  In Fig.~\ref{fig1_2} analogous results are shown for the case when only a nonlinear lattice is present. We consider relatively low currents ensuring stability of the ground state. It is evident from Figs.~\ref{fig1_1} and \ref{fig1_2} that for the low currents assumed hereafter, the bifurcation diagram and the density distribution are nearly the same in the presence and in the absence of the current. However, as we show later, the scattering of the condensate is sensitive to the value of the current.

\section{Bogoliubov-de Gennes spectrum of sound waves}
In the following, we are interested in the sound waves propagating against the background $\psi_{v_0}$ and therefore we look for a solution of Eq.(\ref{GP}) in the form
\begin{eqnarray}
\label{ansatz}
\Psi_{v_0}=e^{-i\mu t+iv_0x}[\psi_{v_0}(x) + u(x)e^{i \Omega t} + \bar{w}(x)e^{-i\bar{\Omega} t}],
\end{eqnarray}
with $|u|,|w| \ll |\psi_{v_0}|.$ At the first order of the perturbative expansion we obtain from the time-independent GP equation the following Bogoliubov - de Gennes equations \cite{de-Gennes}:
\begin{subequations}
\label{sound}
\begin{eqnarray}
 -u_{xx}-2iv_0u_x+ V_1(x)u+V_2(x)w=-\Omega u
\\
-w_{xx}+2iv_0w_x+ V_1(x)w+ \bar{V}_2(x)u=\Omega w,
\end{eqnarray}
\end{subequations}
from which the dispersion relation of BEC sound waves can be determined.
Here $V_1, V_2$ are real and complex potentials, respectively, given by
\begin{eqnarray}
&& V_1(x)= v_0^2+V(x)-\mu+ 2U|\psi_{v_0}|^2, \\
&& V_2(x)= U\psi_{v_0}^2,
\end{eqnarray}
and having, in general, different periods. A negative imaginary part of $\Omega=\omega-i\gamma$ (i.e., a positive $\gamma$) implies the instability of the background $\psi_{v_0}$. The following simple properties of the spectrum can be also derived.
From Eq. (\ref{sound}) it follows that $\Omega=0$ is an eigenvalue with eigenfunction $(u,w) =(\psi_{v_0},-\bar{\psi}_{v_0})$.  More generally, we have that if $\Omega$ is an eigenvalue of (\ref{sound}) with  eigenvector $(u,w)$, then $-\bar\Omega$ is also an eigenvalue corresponding to the eigenvector $(\bar{w},\bar{u})$. It also follows from (\ref{sound}) that $u$ and $w$ are Bloch states characterized, according to Floquet (Bloch) theorem, by the wave vector (crystal momentum) $k$ belonging to the first Brillouin zone, which in our case is  $k\in [-1,1]$. Amplitudes  $u$ and $\bar{w}$ in (\ref{ansatz}) represent excitation amplitudes of forward- and backward-propagating Bloch sound waves with wavevectors $\pm k$.  To identify the bands with positive and negative frequency deviations from the chemical potential we distinguish them by the sign by the subscript as $\omega_{\pm n}(k)$. The symmetry of the sound spectrum is then expressed by the identities
\begin{eqnarray}
\omega_{n}(k)=-\omega_{-n}(-k),\quad \gamma_{n}(k)=\gamma_{-n}(-k).
\end{eqnarray}
\begin{figure}[h]
\epsfig{file=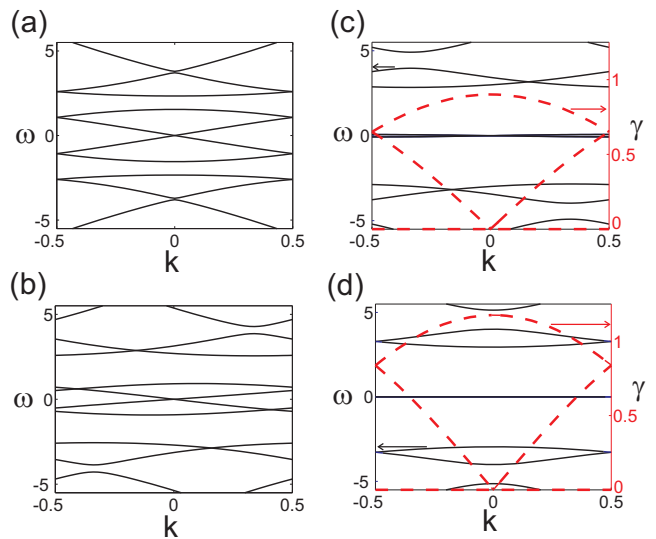,angle=0,width=\columnwidth}
\caption{(Color online) The spectrum of the sound waves in the condensate for $v_0=0$ [panel (a)], $v_0=0.33$ [panel (b)], $v_0=0.66$ [panel (c)], and $v_0=1$ [panel (d)]. The chemical potential of the ground state is $\mu=3$. The solid black and dashed red  curves correspond to the real $\omega$ (left vertical axis) and imaginary $\gamma$ (right vertical axis) parts of $\Omega$.}
\label{fig3}
\vspace{0.1cm}
\end{figure}
Figure \ref{fig3} shows  the sound wave spectrum $\Omega(k)$ for different values of the current velocity $v_0$ of the $\pi-$periodic current state depicted in Fig.~\ref{fig1_1}(b) with constant nonlinearity, i.e., for $U(x)\equiv 1$. It is worth noting  the existence of the threshold velocity $v_{th}\approx 0.6$ such that the background is stable for $v_0<v_{th}$ and unstable for $v_0>v_{th}$. This property is quite expected, being related to the fact, that in the small density limit, the change of the sign of the effective mass of a quasilinear Bloch state implies the change of the stability of the mode~\cite{KS}.

\section{Resonant scattering induced by a moving defect}
A qualitative picture of the resonant scattering between sound waves and moving defects in a periodic condensate in a ring can be obtained by means of perturbation theory. Here we sketch the standard derivation procedure \cite{Landau_add} for the case of the defect with an infinite mass and show how it can be applied for the simplest case of spatially uniform condensate. Assume that we have a linear equation with a right-hand side of the following structure:
\begin{eqnarray}
\hat L  Y =g(x-vt),
\label{add1}
\end{eqnarray}
where $\hat L$ is a linear operator and $g$ is a localized function moving with the velocity $v$. To find the solution of the equation we have to find Green´s function as a solution of the equation
\begin{eqnarray}
\hat L G =\delta(x-vt),
\nonumber
\end{eqnarray}
where $\delta$ is a Dirac $\delta$ function. The solution of Eq.(\ref{add1}) then can be expressed as $Y=\int_{-\infty}^{\infty} g(\xi)G(x-\xi)d\xi$.

One can seek the Green´s function via Laplace transform with respect to time and the expansion over Bloch functions with respect to space. In this representation, Eq.(\ref{add1}) will transform into a linear algebraic equation that can easily be solved. We then need to perform an inverse Laplace transform and integrate over Bloch functions (taking the integral over $k$ for all of the branches of the dispersion characteristic and summarizing over all the branches). It is a well known fact that the stationary solution does not exist if for any $l$, there is a resonance $L(\omega_l=kv,k)=0$ resulting in the poles of the integrand in the inverse Bloch expansion [$L(\omega, k)$ is the image of the operator $\hat L$ in Laplace-Bloch space and $\omega_l$ is the $l$-th branch of the dispersion characteristics of the Bloch waves]. The nonstationary part of the solution can be considered as transitional radiation or as Cherenkov radiation in terms of Bloch functions.

The crucial difference between spatially uniform and periodical problems stems from the fact that in a latter system a moving defect can resonantly couple to a Bloch mode through its higher-order harmonics. It is evident from the band structure associated with the Bloch waves that the resonances of this kind always exist for arbitrary velocity of a defect. This phenomenon is responsible, for example, for the deceleration and decay of moving Bragg solitons \cite{bragg_sol}. In this paper we argue that this resonance results in the scattering of BECs in periodical systems with a moving defect.

To illustrate this on a simple example, we consider a pointlike defect as a small perturbation of an uniform BEC in the absence of the linear $V(x)\equiv 0$ and in the presence of uniform nonlinearity $U(x)\equiv 1$.
In this case, by analogy with Eq.(\ref{ansatz}), we obtain
\begin{eqnarray}
\Psi_{v_0}=e^{-i\mu t+iv_0x}[\sqrt{\mu-v_0^2}+\varphi(x,t)],
\end{eqnarray}
with the linearized sound wave equation  given by
\begin{eqnarray}
i \varphi_t +     \varphi_{xx} +2i v_0\varphi_{x} -(\mu-v_0^2)\left[\varphi+\varphi^{*}\right]=\nonumber\\
V_d(x-v_{d}t) \sqrt{\mu-v_0^2}.\label{linear_2}
\end{eqnarray}
By taking the pointlike defect of the form $V_d(x-v _{d}t)=a_d \delta (x- v_d t)$ we can rewrite the above equation in Fourier space as
\begin{eqnarray}
(\omega - 2 k v_0 - k^2 ) F(k,\omega)-(\mu-v_0^2)\times \nonumber\\ \left[F(k,\omega)+\bar F(-k,-\omega)\right] = a_d\delta(k v_d - \omega)\sqrt{\mu-v_0^2}
\label{linear_2b}
\end{eqnarray}
with $F(k,\omega)= (2\pi)^{-1}\int dx \int  dt e^{-i (k x - \omega t)} f(x,t)$ denoting the Fourier transform with respect to space and time.  The solution of Eq.(\ref{linear_2b}) can be found explicitly as
\begin{equation}
\label{linear_2c}
 F(k,\omega)=-a_d\delta(k v_d - \omega)\sqrt{\mu-v_0^2}\frac{k^2-2 k v_0+\omega}{D(k,\omega)},
\end{equation}
with
\begin{equation}
D(k,\omega)=k^4+2k^2(\mu-v_0^2)-(\omega - 2 k v_0)^2.
\end{equation}
By equating the denominator to zero, i.e.,
\begin{equation}
D(k,\omega)=0
\label{linear_2d}
\end{equation}
one obtains the same  sound-wave spectrum.  From Eq.(\ref{linear_2c}) we see that the value $F(k,\omega)$ is maximal when the resonant condition  $\omega= v_d k$ is satisfied simultaneously with (\ref{linear_2d}), i.e., when the defect velocity matches the phase velocity of a sound excitation.  For $\omega\neq v_d k$ the effect of the defect potential on the sound waves becomes zero.

This result can be extended to quasiuniform condensates with small size spatially extended defects. In this case, we have that for defect off-resonance velocities, the  right-hand side of Eq. (\ref{linear_2b}) becomes negligibly small (e.g., of higher order in the perturbative expansion) due to the fast oscillations in the  Fourier integral reproducing in this way the Bogoliubov-like spectrum  of the unperturbed case. On the contrary, in the  presence of a  resonance,  the defect motion gives rise to secularities (linear growth in time) in the perturbed sound eigenmodes dynamics which may eventually lead to the superfluidity breakdown.

In the case of periodic BECs, however,  the resonant condition   $\omega= v_d k$ must be generalized to include also higher $k$ modes due to  the periodic structure of the sound wave spectrum. Indeed, one can show that,  in this case, the sound waves  are Bloch functions and each Bloch mode with a given $k$ contains all harmonics with $k+n\frac{2\pi}{L_0}$, where $n$ is an integer. So a Bloch wave with the quasimomentum $k$ and frequency $\omega(k)$ can be resonantly excited by harmonics with wave vectors $k+n\frac{2\pi}{L_0}$ and frequency $\omega$. This implies a resonant condition of the form
\begin{equation}
\omega_l(k_{res})=v_d \cdot (k_{res}+n\frac{2\pi}{L_0}),
\label{resonance}
\end{equation}
where $\omega_l$ is the $l$-th branch of the spectrum of linear excitations.

From this simple analysis we draw the following conclusion. For a defect moving inside a periodic BEC trapped in an infinitely long (or very long) ring  there will always be, due to the continuum  (or the almost continuum)  nature of the $k$ space,  values of $k$ for which the condition  (\ref{resonance}) is satisfied. This implies that BEC superfluidity is expected to be broken for any nonzero defect velocity (the time scale on which the breakdown occurs, however, could  depend on defect characteristics). On the other hand, for finite and small rings, the discreteness of the $k$ space gives rise to the possibility to avoid some (or all) of the resonances leading to quasisuperfluidity (superfluidity) in the presence of a moving defect. We shall investigate these possibilities in more detail in the next sections by means of direct numerical integrations of the GPE.

\section{Long BEC ring with moving defect}
To investigate the possibility of the superfluidity breakdown via resonant scattering, we shall perform direct
numerical integrations of the GPE (\ref{GP}) on a very large ring (provided that the spectrum of the sound waves is quasi-continuous, the results obtained for the infinite system remain valid for these systems too). In particular, in this section we check predictions of the previous section by direct comparison of the Fourier spectra of numerically obtained
GPE periodic BEC solutions
\begin{eqnarray}
\label{spectrum}
S(\tilde k,t)=\left|\int_0^L \Psi (x,t) e^{i\tilde k x} dx\right|
\end{eqnarray}
with the resonance condition in Eq. (\ref{resonance}). In all the following simulations the moving defect is taken of Gaussian form
\begin{eqnarray}
\label{defect}
V_d(x)=a_d \exp\left( \frac{(x-v_d t)^2}{\ell_d^2} \right).
\end{eqnarray}

\begin{figure}[h]
\epsfig{file=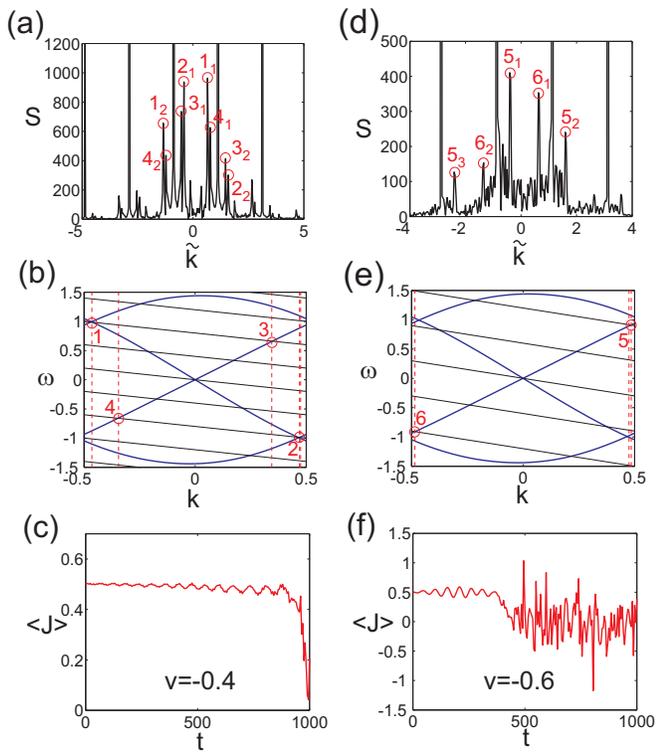,angle=0,width=\columnwidth}
\caption{(Color online) The upper row shows Fourier spectra (\ref{spectrum}) of the condensate order parameter function obtained from direct numerical simulation of Eq. (\ref{GP}). The spectra were calculated at time $t=100$. Graphical solutions of the resonance equation (\ref{resonance}) are shown in the middle row. For a graphical visualization  of the resonance condition  (\ref{resonance}) (red circles) the defect  characteristics  $\omega=v_d k$ are indicated by black lines the in middle  panels. The lower row illustrates the dynamics of the averaged current $<J>$ calculated by formula (\ref{J}). All the results are shown for the nonlinear potential (\ref{potentialU}) with $\kappa=0.25.$, the chemical potential of the background $\mu=-2$, the average velocity $v_0=0.125$, and the  defect velocities $v_d=-0.4$ (left column) and $v_d=-0.6$ (right column). The defect is described by (\ref{defect}) with the parameters $a_d=0.15$, $\ell_d=0.8$, and the total width of the system is $64 \pi$.}
\label{fig4_2}
\vspace{0.1cm}
\end{figure}

Results obtained for the case of  a pure nonlinear OL (e.g., in the absence of the linear potential)
are reported in Fig.~\ref{fig4_2}. In the top left and right panels of this figure we show the GPE Fourier spectrum of the solution at the fixed time $t=100$ for two different values of the defect velocity $v_d=-0.4$ and $v_d=-0.6$, respectively. In the corresponding left and right middle panels we depict the lower branches in the Bogoliubov-de Gennes spectrum of the sound waves. Resonances in the spectrum  are denoted by  red circles (numbered $1$-$6$)  in (\ref{resonance}), and for a graphical visualization of the resonance condition we have reported the defect  characteristic lines  $\omega=v_d k$ in the first Brillouin zone with black lines.

The Fourier modes $\tilde k$ (red circles) in the numerical GPE  spectrum which correspond  to the same  resonant mode  $\omega(k)$ in the  Bloch sound wave spectrum have been marked with subindex numbering of the $n$-th Fourier harmonics. Note that to each resonant Bloch mode there are, in principle, an infinite number of equidistant Fourier harmonics, with the spacing between them  equal to the width of the Brillouin zone, with  $\tilde k$ connected to the quasimomentum $k$ by the relation $\tilde k= k + v_0+ n K $, where $n$ is an integer and $K$ is the width of the Brillouin  zone ($K=1$ for the considered case). The dashed red vertical lines in the middle panels [Figs.\ref{fig4_2}(b) and \ref{fig4_2}(e)]  show the positions of the resonances extracted from numerical simulations using the different spectral lines shown in the corresponding  top panels  [Figs.\ref{fig4_2}(a) and \ref{fig4_2}(d)]. From this we see that there is a perfect agreement between the resonances predicted by Eq. (\ref{resonance}) and the ones obtained from numerical GPE integrations.

Superfluidity  breakdown induced by resonant scattering with a moving defect can be detected by means of the averaged current,
\begin{eqnarray}
\label{J}
\langle J\rangle =\frac 1 L \int_0^L J dx,\qquad  J =\frac{1}{2i}\left(\psi^{*}  \psi_x -\psi  \psi_x^{*}\right),
\end{eqnarray}
 since it is clear that a significant deviation from an uniform time behavior of $\langle J\rangle$ with a drastic drop of the current in time, signals a strong perturbation of the density and a loss of coherence of the condensate.

\begin{figure}[h]
\epsfig{file=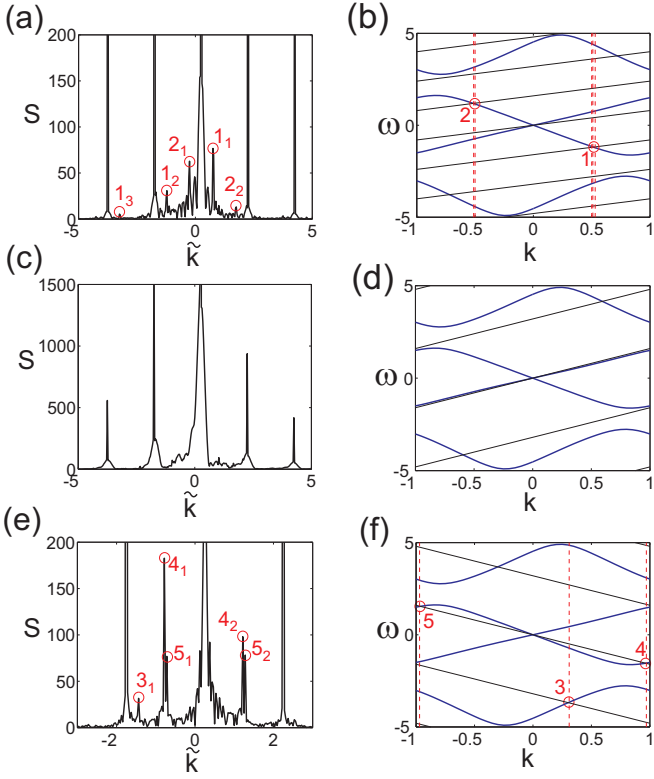,angle=0,width=\columnwidth}
\caption{(Color online)  The left column shows Fourier spectra (\ref{spectrum}) of the order parameter function obtained from direct numerical simulations of the GPE (\ref{GP}). The ground state is characterized by chemical potential $\mu=3$ and velocity $v_0=0.25$; the parameters of the linear potential are $V_0=15$, $\ell =0.15$, and $x_s=0$. In the right column, a graphical solution of the resonance equation (\ref{resonance}) is shown. For a graphical visualization  of the resonance condition  (\ref{resonance}) (red circles) the defect  characteristics  $\omega=v_d k$ are indicated by black lines in the right column. The velocity of moving defect is  $v_d=0.8$ [panels (a) and (b)], $v_d=1.6$ [panels (c) and (d)], and $v_d=-1.6$ [panels (e) and (f)]. The spectra shown in panels (a), (c), and (e)  are calculated at $t=50$, $t=60$, and at $t=100$, respectively. The defect is described by (\ref{defect}) with the parameters $a_d=0.15$, $\ell_d=0.8$, the total width of the system is $64 \pi$.
}
\label{fig4}
\vspace{0.1cm}
\end{figure}

This is shown in Figs.~\ref{fig4_2}(c) and \ref{fig4_2}(f) where  the time dependence  of $\langle J\rangle $ is  presented for the  same values of the defect velocity of Figs.~\ref{fig4_2}(a) and \ref{fig4_2}(b) and then \ref{fig4_2}(d) and \ref{fig4_2}(e), respectively. In both cases, one can see that at the early stage the time evolution of the current exhibits a regular oscillating behavior  around its unperturbed value and, after some time, [$t\approx 900$ for Fig.~\ref{fig4_2}(c) and $t\approx 400$ for Fig.~\ref{fig4_2}(f)], the current drops down significantly and begins to behave chaotically [see Fig.~\ref{fig4_2}(f)]. Notice that for higher  defect velocity the breakdown occurs earlier due to the fact that, in this case, the  perturbation exerted by the defect on the condensate is stronger.

\begin{figure}[h]
\epsfig{file=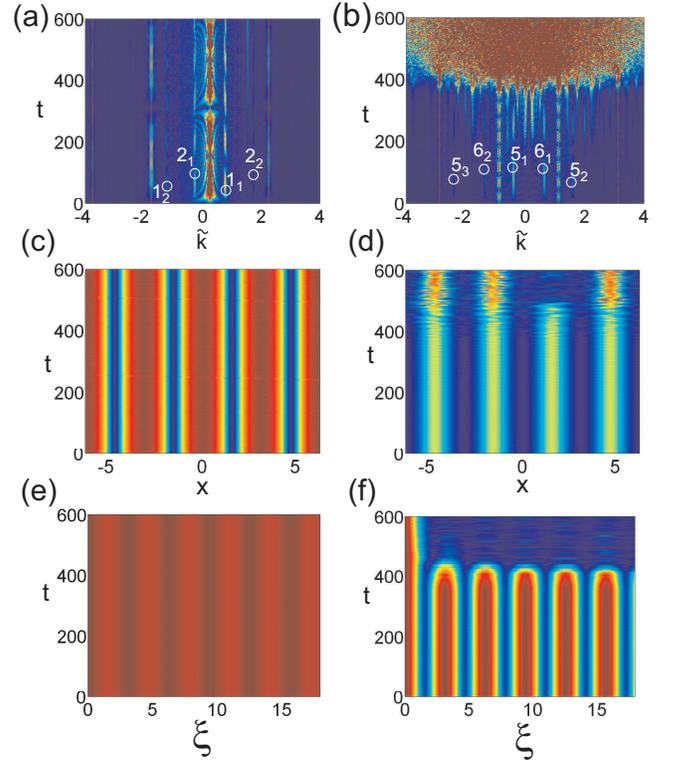,angle=0,width=\columnwidth}
\caption{(Color online) The dynamics of the Fourier spectra (\ref{spectrum}) for the cases of linear [panel (a)] and nonlinear [panel (b)] lattices  (upper row). In the middle row, the dynamics of the condensate density is shown in panels (c) and (d) for the cases of linear and nonlinear lattices, correspondingly. The bottom row [panels (e) and (f)] shows the dynamics of the correlation function (\ref{corr}) for the same cases. For panels (a), (c), and (e) the parameters are the same as for panels (a) and (b) of Fig.~\ref{fig4}; the parameters for panels (b), (d), and (f) are the same as for panels (d), (e), and (f) of Fig.~\ref{fig4_2}. Circles with numbers in panels (a) and (b) denote the resonant lines shown in panel (a) of Fig.~\ref{fig4} and  panel (d) of Fig.~\ref{fig4_2}, respectively.
}
\label{fig4_3}
\vspace{0.1cm}
\end{figure}
A similar analysis performed for the case of a pure linear OL (e.g., in the presence of a spatially uniform nonlinearity along the ring) is reported in Fig.~\ref{fig4}. The left column panels of this figure show the spectra obtained from the numerical modeling of GPE (\ref{GP}) for different velocities of the defect. The Bogoliubov-de Gennes spectra of the sound waves and the dispersion characteristics of the defect are shown in the right column panels. The resonances between the defect and the Bloch modes of the condensate are marked by red circles $1$-$5$ in the right column.
The Fourier harmonics corresponding to the $n$-th resonant Bloch mode are marked in the spectra shown in the left column by the subindex numbering the harmonic. The dashed red vertical lines in the right column show the positions of the resonances extracted from direct numerical simulations using different spectral lines shown in the left column. One can see that the predicted and the measured positions of the harmonics are in an excellent agreement. The discrepancy can be explained by the width of the corresponding spectral lines.

It is worth mentioning here that a special case is shown in Figs.~\ref{fig4}(c) and \ref{fig4}(d) where we observe peaks that  not well resolved  with widened spectral lines. This behavior stems from the dispersion characteristics of the defect that is nearly tangential to that of the Bloch waves for the lower branches  while the resonances associated with the  higher modes are exponentially weak and, thus, not visible. We also remark that, as for Fig.~\ref{fig4_2}, the dispersion of the Bloch waves are calculated for the functions given by formula (\ref{ansatz}) and, therefore, the wave vectors $\tilde k$ of the Fourier harmonics of a Bloch mode are connected to the quasimomentum of the Bloch wave by the relation $\tilde k= k + v_0+n K$, where $n$ is an integer and $K$ is the width of the Brillouin  zone equal to $K=2$ for the considered case.

Quite interestingly, and in contrast with the previously considered case of a pure nonlinear OL, the effect of the resonant scattering on the condensate stability seems to be negligible. In Fig.~\ref{fig4_3} the differences between the pure linear and pure nonlinear OLs cases with respect to the superfluidity breakdown induced by a moving defect are further investigated. The upper row panels shows the dynamics of the power spectrum obtained from GPE numerical integrations for pure  linear (left panel) and pure nonlinear (right panel). We see that at  the early stages of the evolution  the intensity of the resonant spectral lines  oscillates in  time. During the oscillations the spectral lines corresponding to the resonant scattering become wider due to nonlinear effects and new lines appear in the spectrum. In the case of pure nonlinear OLs, however,  after a certain time  the perturbation begins to grow and the  ground state is destroyed [this occurs at $t\approx 400$, as one can see from Figs.~\ref{fig4_3}(b), \ref{fig4_3}(d), and \ref{fig4_3}(f)].

To inspect the superfluidity of the condensate, we also evaluate, in addition to the condensate density, a correlation function which reads
\begin{eqnarray}
\label{corr}
\Gamma(\xi,t)=\left| \frac{1}{L}\int_0^L \psi(x,t)\psi^*(x-\xi,t) dx  \right|.
\end{eqnarray}

The correlation function proves to be a more sensitive tool for detecting the loss of coherence of the condensate. In the case of a pure nonlinear OL the superfluidity breakdown is evident from Figs.~\ref{fig4_3}(d) and \ref{fig4_3}(f), where  the dynamics of the BEC density  and of the correlation function are presented, respectively. One can see that, in this case, all the characteristics indicate the loss of superfluidity. At the same time, for a pure linear optical lattice, not only does the spectrum or the density of the condensate remain practically the same, but also the correlation function shows the condensate is in a superfluid state.

Note that the superfluidity breakdown in the Fourier spectra corresponds to a transition from  a quasidiscrete spectrum to a quasicontinuum one [cf. Fig.~\ref{fig4_3}(b)]. This behavior is in contrast to that observed in Figs.~\ref{fig4_3}(c) and \ref{fig4_3}(e) for the case of pure linear OLs for which, in spite of the presence of the resonances in the spectrum, the corresponding dynamics remains regular in time.

The explanation of the  noneffectiveness of the resonant scattering in the case of linear OLs can be assigned to the nonlinear nature of the phenomenon. To this end, we note that the resonant scattering as discussed in the previous section is a purely linear effect. On the other hand the system is intrinsically nonlinear and a linear resonance can induce a more complicated (nonlinear) dynamics. In the case of the nonlinear OLs the condensate exists in the form of relatively weakly interacting droplets (for the given chemical potential $\mu$) and strong depletion of the condensate in the areas between the droplets enhance the coupling between the defect and the resonant Bloch waves, making their excitation more effective. At the same time weak coupling between the condensate droplets impairs the stability of the condensate. However, in the considered case of the linear lattice, the depletion of the condensate is relatively weak and, as a result, this condensate is spatially very stable and cannot be destroyed via the linear resonant scattering.

We want to emphasize that we do not claim that the condensates in the nonlinear lattices are always less stable against the perturbations introduced by moving defects. One can expect that under special circumstances and for different sets of the parameters, in particular for the condensate with strongly depleted areas, the coherent state of the condensate could be  destroyed  during quite short time by means of resonant scattering in the case of pure linear OLs as well. This problem is discussed in more detail in the next section.

\section{BEC superfluidity in mesoscopic size rings with linear OL}

The main aim of the section is twofold. The first aim is to consider the breakdown of the BEC placed in a linear OL and the second one is to answer the question "How does the discreetness of the sound wave spectrum affect the dynamics of the condensate perturbed by a moving defect?" To answer this question, we concentrate on the particular setting of rings of finite (not very large) sizes  and deep pure linear OLs. Let us note here that we do not assume that the number of particles in the condensate is low and so our parameters do not contradict the applicability of the Gross-Pitaevskii equation. The only thing which is required is that the characteristic width of the Cherenkov resonance is considerably smaller than the distance between the harmonics in the spectrum of the sound waves of the condensate.

The reasons for these choices are the following. For rings of small size, the set of the eigenmodes in Bogoliubov-de Gennes spectrum can be very discrete due to discrete nature of the $k$ space, $k=\frac{2\pi n}{L}$ with  $n$ an integer. In this case, one  can expect that a mode can be resonantly excited only if the synchronism between defect and sound waves  is exactly fulfilled for the allowed values of $k$, e.g., one can use  the discreteness in the spectrum to  manage the resonances. This can be done by properly adjusting the defect velocity to miss (or to match) a resonance, leading to the destruction (or preservation) of the BEC superfluidity. This idea is of general validity, e.g., valid for linear and  nonlinear OLs.

According to our discussion, the destruction of the condensate for typical situations requires that the intensity of the resonant harmonic grows beyond a certain threshold for the nonlinearity to become important. This situation can be realized more easier in the presence of a very deep  OL when the condensate is fragmented in an  array of  weakly interacting droplets and the efficiency of the excitation of high order modes is enhanced. Another factor making it easier to observe the destruction of the coherent state of the condensate in a deep OL is that the superfluid property of the whole array in this case is granted by the coherence kept by the condensate in the weakly overlapping regions linking the adjacent droplets. These links, however, are weak, which means that the stability of coherence between the neighboring droplets is impaired. In the following we shall investigate this on the specific example of  a condensate in a ring of length $L=8 \pi$ resting with respect to a pure linear OL  of strength $V_0=100$.

\begin{figure}[h]
\epsfig{file=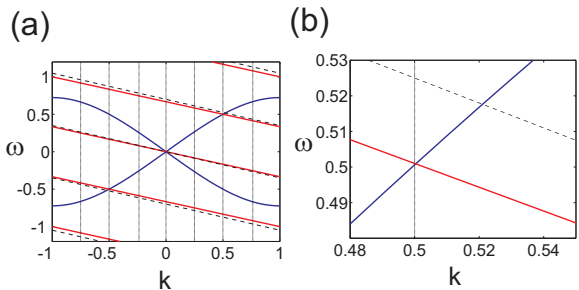,angle=0,width=\columnwidth}
\caption{(Color online) The resonance condition for the velocities $v=-0.334$ (solid red lines) and $v_d=-0.35$ (dashed black lines). Panel (a) shows the whole lowest branch of the dispersion characteristics; panel (b) shows a close-up of the resonance. Vertical dotted lines show the allowed values of $k$ for the trap length $L=8 \pi$. The parameters are $V_0=100$, $w=0.15$, $\mu=5$, $v_0=0$.}
\label{fig7}
\vspace{0.1cm}
\end{figure}

A typical spectrum of the sound wave excitation for a ring of length $L=8\pi$ is presented in Fig.~\ref{fig7} (vertical black lines denote the allowed discrete values of $k$ in this case).  Using the resonant condition, we can find that the mode with $k=0.5$ in the sound wave spectrum [see Fig.~\ref{fig7}(a)] can be excited if the velocity of the obstacle is $v_d=-0.334$. If the velocity of the obstacle deviates from this value, then the mode becomes detuned from the resonance. This can be seen more clearly from Fig.~\ref{fig7}(b), where the resonances for the obstacles moving with the resonant velocities $v_d=-0.334$ and  $v_d=-0.35$ are shown on  enlarged scales.

\begin{figure}[h]
\epsfig{file=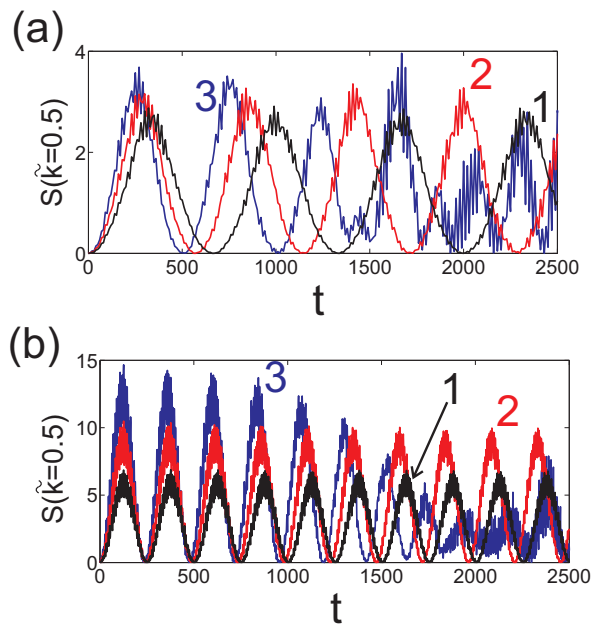,angle=0,width=\columnwidth}
\caption{(Color online) Panel (a) shows temporal dynamics of the $S(\tilde{k})$ (in arbitrary units) with $\tilde k=0.5$ when the defect is moving at the velocity $v=-0.334$ and the resonant condition is held for the allowed value of $k$ (for $k=0.5$). The black line marked as $1$ is for the case when $a_d=0.4$, and the red and blue lines marked as $2$ and $3$ are for $a_d=0.5$ and $a_d=0.6$ correspondingly. Panel (b) shows the same but for the defect moving at $v_d=-0.35$. The width of the moving defect is $\ell_d=0.125$. The other parameters are the same as in Fig.~\ref{fig7}.}
\label{fig8}
\vspace{0.1cm}
\end{figure}

In Fig.~\ref{fig8} we show  the temporal dependence of the intensity of the Fourier harmonics $S(\tilde{k})$ (15) at  $k=0.5$  for different heights of the potential modeling a moving defect in the case of  resonant [Figure \ref{fig8}(a)] and off-resonant [Figure \ref{fig8}(b)] velocity.
Figure \ref{fig8}(a) shows that for the resonant velocity $v_d=-0.334$ the dynamics is quasiperiodic and the amplitude of the oscillations does not depend significantly on the strength of the scattering potential. However, the period of the oscillations depends on the strength of the potential significantly. Let us note here that assuming that $v_d=-0.334$ provides nearly exact synchronism we should expect that in a linear regime the intensity of the resonant harmonic should grow quadratically in time. It can be checked that for the velocity $v_d=-0.334$ the period becomes $T_l \approx 6000$ (at exact resonance the period becomes infinite). From numerical simulations performed for $v_d=-0.334$ we can see  that the period of the oscillations begins to differ markedly from the one predicted by the linear theory. From this  we conclude that the nature of the oscillations in the case of the resonant excitation must be strongly nonlinear.

In the case of off-resonant velocity $v_d=-0.35$ the amplitude of the Fourier harmonics oscillate in time; however, the amplitude of the oscillations depends on the strength of the potential much more profoundly than in the case of $v_d=-0.334$; see Fig.~\ref{fig8}(b). We also observe that the period of the oscillations slightly depends on their amplitude but if the height of the scattering potential decreases, then the period of the oscillations reaches a constant value which can easily be found from the linear theory and reads $T_{l}=\max \frac{2\pi L}{|\omega_r L -2\pi v_d m|}$, where $\omega_r$ is the resonant frequency of the mode and $m$ is an integer.

For the chosen parameters $L=8\pi$ and $\omega_r \approx 0.5$ the period of linear oscillations is $T_l=251.3$ obtained for $m=6$. This means that the oscillations with the longest period are excited by the sixth harmonics of the driving force (produced by a moving defect). From our numerical simulations we obtain the period of the oscillation $T=245.7$ for $a_d=0.5$ and $T=250.5$ for $a_d=0.4$. This shows that for smaller  amplitudes of the excited harmonics the period of the oscillation tends to its linear value $T_{l}$.

Similarly to the pure linear OL case investigated in the previous section, it is difficult to detect any superfluidity breakdown  in terms of the  evolutions of Fourier spectrum or condensate density distribution. On the other hand, as we also noted before, the correlation function is a more accurate tool to measure the coherence of the condensate. For a periodic BEC in the ring to be in  superfluid state the correlation function must be periodic with the same periodicity of the OL and  behave regularly in time (it should not grow or decay). However, when the condensate becomes strongly perturbed by the defect motion, e.g., when coherence and superfluidity in linking regions between droplets are destroyed,  the correlation function cannot be periodic  with the period of the  OL (period extend to the whole ring length  $L$) and for $\xi$ greater than the OL period the amplitude of the correlation function will decrease significantly.

\begin{figure}[h]
\epsfig{file=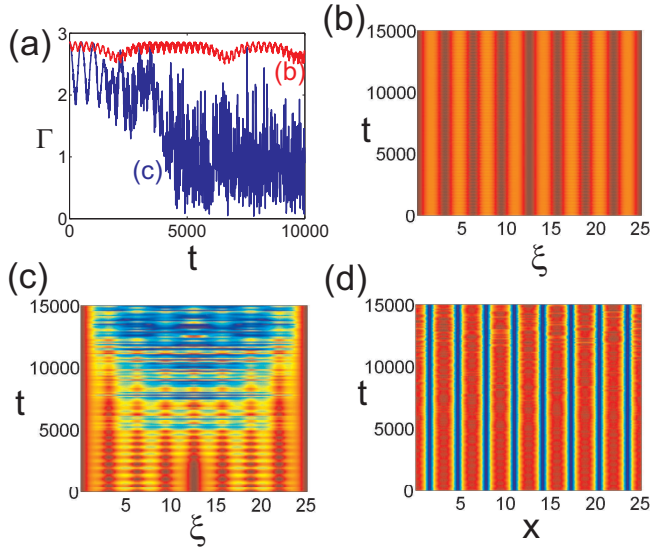,angle=0,width=\columnwidth}
\caption{(Color online) Panel (a) shows temporal dynamics of the correlation function (\ref{corr})
%$\Gamma(\xi)=\left| \frac{1}{L}\int_0^L \psi(x)\cdot \psi^*(x-\xi) dx \right|$
at $\xi=6.4$, the blue line marked as (c) is for $v_d=-0.334$ and the red line marked as (b) is for $v_d=-0.35$. Panels (b) and (c) shows temporal evolutions of correlation function $\Gamma$ for $v_d=-0.35$ and $v_d=-0.334$ correspondingly. Panel (d) shows the dynamics of the density of the condensate for $v_d=-0.334$.  The height of the potential modeling the defect is $a_d=0.5$.  The other parameters are the same as in Fig.~\ref{fig8}.}
\label{fig9}
\vspace{0.1cm}
\end{figure}

This is exactly what we obtain for the correlation function from GPE numerical integrations [see Figs.~\ref{fig9}(a)-\ref{fig9}(c)] for the resonant and nonresonant cases. In particular, in Fig.\ref{fig9}(a) the temporal dependence of the correlation function at $\xi=6.4$ is depicted. One can see that while for the resonant velocity [see blue line also marked as (c) in the panel] the correlation function does a few oscillations and then decays to lower values, the red line  corresponding to the off-resonant case [also marked as (b) in the panel] remains approximately constant. Behavior of the correlation function $\Gamma(t, \xi)$ is shown in Figs.~\ref{fig9}(c) and \ref{fig9}(b) for the resonant and off-resonant cases, respectively.  From Fig.~\ref{fig9}(c) one can also clearly see the loss of periodicity and the decay of the correlation function  at $t\approx 5000$ signaling the superfluidity breakdown of the condensate.

Let us now describe in more detail how the destruction of the superfluid state occurs in this case.  We notice that even if the condensate is strongly perturbed by a defect and is no longer in a coherent ground state the distribution of the density of the condensate does not necessarily  undergo  drastic changes as in the case of nonlinear lattices. Due to the repulsive nonlinear interaction the density will be more or less evenly distributed outside the repelling areas of the external linear potential (in these areas the condensate will be depleted). However,  the coherence of the weakly coupled neighboring droplets can be destroyed by the moving defect.

From these results we  conclude that for the considered system  the destruction of the superfluid state occurs when the defect velocity can match the resonance condition  between the moving defect and an eigenmode.
It is also worth noting that this scenario of the breaking of the coherent state of the condensate differs very much from the scenario considered in the previous section for the nonlinear lattice for which the superfluidity  breakdown can be detected in terms of BEC density as well. From Fig.~\ref{fig9}(d), indeed,  we see that for  the resonant velocity $v_d=-0.334$ the density  is practically uniform in time and  still resembles the initial density for $t>5000$ when the phase correlation starts to disappear.

In the conclusion of this section we would like to mention that, as the strength of the  OL is decreased,  the superfluidity will be lost on very long time scales  (it could be longer than the lifetime of the condensate) and the  phenomenon becomes not detectable. The case of very deep linear OLs, however, is certainly  experimentally accessible and  we hope that our  predictions about this interesting phenomenon  could be be confirmed  in the future.

Finally, an interesting question to pose is if the above resonant scattering mechanism  for the BEC superfluidity breakdown  could be  effective  also in a three-dimensional (3D)  setting. In this respect we remark  that the presence of competitive mechanisms typical of the 3D case, like  collapse,  delocalizing thresholds, existence of vortices, etc., makes it impossible to extrapolate our results to the 3D case and to provide an answer to the above question; additional extensive work, beyond the scope of the present paper, would be  required.

\section{Conclusions}

In this paper we performed analytical  and numerical investigations of the quasisuperfluidity properties of periodic BEC in quasi-1D rings with linear and nonlinear  OLs and in the presence of a moving  defect. We have shown that the resonant scattering of sound Bloch waves with the defect profoundly affect the BEC superfluidity and in spite of the presence of a spectrum of the excitations of Bogoliubov type, a  moving defect  can lead to the breakdown of BEC superfluidity, independently of the value of the defect velocity.

In particular, we have shown the existence of  an instability which arises from the resonance in the defect-sound waves dynamics (resonant scattering) which becomes evident in topological geometries. This instability, although of  dynamical type, differs from  the usual  dynamical (e.g., modulational) instability of periodic BEC systems because it does not lead to the creation of solitons. We also demonstrated that for weak periodic potentials the superfluidity breakdown may occur on a long time scale (quasisuperfluidity) and  the breakdown process can be accelerated by increasing the strength of the OL. Finally, we have shown that when the size of the ring becomes small enough to imply the discreteness of the reciprocal space, it is possible to avoid the resonant scattering and to restore quasisuperfluidity or superfluidity behavior. We remark that the observed resonances open a way to measure the spectral properties of the sound waves in BECs by varying the velocity of a probing defect and analyzing the respective scattering data.  For this the study of the correlation function of the condensate is a more effective tool (than the density) to detect the loss of the phase coherence.

The obtained results for pure linear and pure nonlinear OLs  open a new scenario  for the BEC superfluidity breakdown in topological quasi-1D settings  in the presence of a moving defect. We also remark that the developed theory can be directly applied to the polariton BEC in periodical annular 1D semiconductor microcavities.
\bigskip

\section*{Acknowledgements}
A.V.Y and V.V.K acknowledge support of the FCT (Portugal) under Grants No. PTDC/FIS/112624/2009, and PEst-OE/FIS/UI0618/2011. Y.V.B. acknowledges the support of FCT  (Portugal) under Grant No. PEst-C/FIS/UI0607/2011.
A.V.Y and V.K. were  supported by  the bilateral project.  M.S. acknowledges
partial support from the Ministero dell' Istruzione,
dell' Universit\'a e della Ricerca (MIUR) through a {\it Programma
di Ricerca Scientifica di Rilevante Interesse
Nazionale} (PRIN)-2010 initiative.

\end{document}